\documentclass[screen]{acmart}

\AtBeginDocument{%
  }

\setcopyright{acmlicensed}
\copyrightyear{2025}
\acmYear{2025}
\acmDOI{XXXXXXX.XXXXXXX}

\begin{document}

\title{FAIR Ecosystems for Science at Scale}

\author{Sean R. Wilkinson}
\affiliation{%
  \institution{Oak Ridge Leadership Computing Facility}
  \city{Oak Ridge}
  \state{Tennessee}
  \country{USA}
}
\email{wilkinsonsr@ornl.gov}
\orcid{0000-0002-1443-7479}

\author{Patrick Widener}
\affiliation{%
  \institution{Oak Ridge Leadership Computing Facility}
  \city{Oak Ridge}
  \state{Tennessee}
  \country{USA}
}
\email{widenerpm@ornl.gov}
\orcid{0000-0002-5882-0816}

%\renewcommand{\shortauthors}{Wilkinson et al.}

\begin{abstract}
High Performance Computing (HPC) centers provide resources to users who require greater scale to ``get science done''. They deploy infrastructure with singular hardware architectures, cutting-edge software environments, and stricter security measures as compared with users' own resources. As a result, users often create and configure digital artifacts in ways that are specialized for the unique infrastructure at a given HPC center. Each user of that center will face similar challenges as they develop specialized solutions to take full advantages of the center's resources, potentially resulting in significant duplication of effort. Much duplicated effort could be avoided, however, if users of these centers found it easier to discover others' solutions and artifacts as well as share their own.

The FAIR principles address this problem by presenting guidelines focused around metadata practices to be implemented by vaguely defined ``communities''; in practice, these tend to gather by domain (e.g. bioinformatics, geosciences, agriculture). Domain-based communities can unfortunately end up functioning as silos that tend both to inhibit sharing of solutions and best practices as well as to encourage fragile and unsustainable improvised solutions in the absence of best-practice guidance. We propose that these communities pursuing ``science at scale'' be nurtured both individually and collectively by HPC centers so that users can take advantage of shared challenges across disciplines and potentially across HPC centers. We describe an architecture based on the EOSC-Life FAIR Workflows Collaboratory, specialized for use with and inside HPC centers such as the Oak Ridge Leadership Computing Facility (OLCF), and we speculate on user incentives to encourage adoption. We note that a focus on FAIR workflow components rather than FAIR workflows is more likely to benefit the users of HPC centers.
\end{abstract}





\maketitle

\section{Introduction}
% INTRODUCTION
%

%\subsection{The FAIR Principles}

The FAIR principles for making digital objects Findable, Accessible, Interoperable, and Reusable were first established to improve the management and stewardship of data \cite{wilkinson2016}. FAIR's focus on metadata has since found broad application beyond data to research software \cite{barker2022}, artificial intelligence (AI) and machine learning (ML) models \cite{huerta2023}, and computational workflows \cite{wilkinson2025}.
Following the FAIR principles helps make research artifacts more easily discoverable, shareable, and [re]usable. These principles improve collaboration, transparency, and efficiency in scientific research by ensuring that artifacts like code and data are well-documented, allowing them to be reused effectively across different studies and disciplines both by humans and machines. There may be challenges in implementing the FAIR principles \cite{wilkinson_f_2022}, but they do play an important role in facilitating reproducibility, accelerating discoveries, and supporting long-term stewardship of research artifacts. Their adoption helps researchers maximize the impact of their artifacts while fostering open science and innovation.

%\subsection{FAIR communities}

The FAIR principles focus on metadata practices to be implemented by vaguely defined ``communities''; in practice, these tend to gather by domain (e.g. bioinformatics, geosciences, and agriculture). The efforts have improved sharing and reusing artifacts within their respective disciplines, but these domain-based communities end up functioning as silos. This hampers sharing and reusing artifacts as well as the adoption of the FAIR principles themselves across computational research areas.

For example, the bioinformatics community has made significant strides in developing FAIR repositories such as the European Nucleotide Archive \cite{yuan2023}. Similarly, the geoscience community has developed metadata-rich repositories such as EOSDIS \cite{ramapriyan2020}, which provides comprehensive access to satellite and remote sensing data while supporting research in areas such as climate change, natural disasters, and ecosystem monitoring. However, interoperability between these repositories remains limited, reducing the potential for cross-domain collaboration.

%\subsection{Challenges at HPC centers}

High Performance Computing (HPC) centers present additional challenges not present in commodity computing environments that must be addressed when implementing FAIR principles. In order to provide resources to users who require greater scale to ``get science done'', HPC centers frequently deploy infrastructure with exotic hardware architectures, cutting-edge software environments, and stricter security measures as compared with users' own resources \cite{antypas2021}. Variability in system architectures, ranging from CPU-based clusters to GPU- and TPU-accelerated platforms, complicates software portability and reproducibility. Furthermore, interdisciplinary and inter-project collaboration within and across HPC centers remains challenging due to security constraints, data governance policies, and technical barriers. As a result, users often create, configure, and customize digital artifacts in ways that are specialized for the unique infrastructure at a given HPC center. This means that the research artifacts produced at HPC centers have limited potential for reuse by others unless they are shared in ways that are discoverable by other users of those HPC centers.

%\subsection{Existing US DOE efforts}

Several efforts in the United States (US) Department of Energy (DOE) share the common goal of advancing scientific research by providing cutting-edge computational, data, and networking resources to support interdisciplinary collaboration: the Integrated Research Infrastructure (IRI) \cite{miller2023}, the Framework for Accelerating Science and Scientific Training (FASST) \cite{fasst2025}, and the National Artificial Intelligence Research Resource (NAIRR) \cite{nairr2025}. These initiatives focus on improving access to high-performance computing (HPC), data management, and security frameworks to foster cross-disciplinary research and enable more efficient, data-driven scientific discoveries. They emphasize the importance of open science principles and FAIR, ensuring that research data and tools are accessible and reusable. These efforts are more concerned with enabling the autonomous laboratories of the future, however, rather than the human users of HPC facilities.

%\subsection{EOSC-Life FAIR Workflows Collaboratory}
The EOSC-Life FAIR Workflows Collaboratory is a European initiative aimed at enhancing the FAIRness of scientific workflows in life sciences \cite{goble_implementing_2021}. It provides a collaborative platform, including WorkflowHub \cite{gustafsson2024}, that enables researchers to share, develop, and execute workflows across various life science domains, ensuring that data, tools, and computational resources are well-documented and standardized for broader use. It leverages the European Open Science Cloud (EOSC) to foster collaboration between institutions and projects, promote transparency, and improve the efficiency of research through the integration of diverse data sources and computational tools. The initiative is crucial for supporting reproducibility, advancing data-driven discoveries, and accelerating innovation in European life sciences.

In this paper, we argue that users could increase impactful scientific output if HPC centers enhanced their unique environments with a FAIR-based strategy similar to that pursued by EOSC-Life. We observe that a focus on FAIR workflow components may be more impactful than focusing on FAIR for entire workflows. We also propose a general architecture and discuss how to encourage adoption by users.

\section{Ecosystem architecture}
% ARCHITECTURE

FAIR ecosystems for HPC are better served by focusing on the sharing, discovery, and reuse of workflow components rather than entire workflows.
This is partly because HPC workflows are often short-lived due to being customized for HPC resources with short design lifetimes (due to the rapid pace of hardware and software innovation). For example, the flagship machines at centers like OLCF are designed for a lifespan of 5 years; after that, executing the same workflows on a machine's successor is almost guaranteed to fail. Having witnessed this process repeatedly, we have noticed that workflow components themselves are less likely to break, but they will have to be re-plumbed to port the workflow to the new machine. Another reason for focusing on components is that, in our experience, it is extremely rare to locate an existing off-the-shelf workflow that exactly matches your needs. Users at HPC centers come from a variety of scientific disciplines, so it is unlikely that any particular workflow will match the needs of other users exactly. Because these users are all solving challenges in trying to scale using the same HPC resources, however, it is much more likely that solutions to separable pieces of problems will be reusable across disciplines, supporting a component-based approach. For these reasons we propose to focus on applying FAIR to workflow components, making sure they are well-described by rich metadata. This strategy requires a combination of repositories, registries, computing infrastructure, authentication/authorization, and metadata standards.

%%%
\begin{figure}[htbp!]
\centering
\includegraphics[width=.9\textwidth]{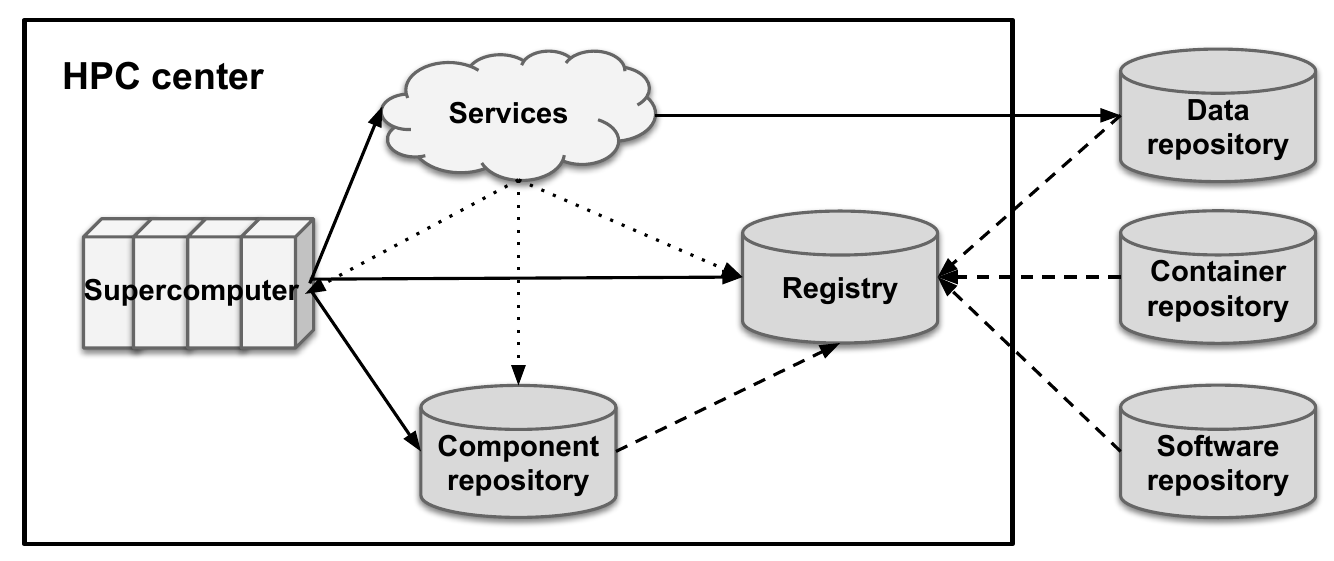}
\caption{\label{fig:diagram}Illustrated examples of a FAIR ecosystem in action at an HPC center. Dashed lines (\textendash\textendash) show the registry's records tracking components wherever they are. Dotted lines ($\cdots$) show services in an on-prem cloud monitoring the registry and component repository for updates and providing support such as databases to supercomputer jobs. Solid lines (\textemdash) show that the supercomputer can consult the registry to locate components and request the service infrastructure to retrieve external artifacts if needed.}
\Description{}
\end{figure}
%%%

\noindent\textbf{Repositories}, roughly speaking, are places where \textit{things} are stored; this is in contrast to registries (see below), which are where \textit{records} of things are stored. Workflow components are what workflows are made of, which can include various forms of data (e.g. simulation output, trained AI/ML models, and provenance) and software (e.g. scripts, programs, and computational workflows). Having a robust repository system aligns with IRI-style distributed, collaborative science and its aims for creating a seamless, interoperable research infrastructure across DOE labs.

The FAIR principles do not imply openness, but we also recommend following open science principles \cite{nasem2018} when possible by enabling integration with externally hosted public repositories (e.g. Conda, Spack, Docker Hub, TensorFlow Hub). Security or scale concerns may not allow for open dissemination or integration, however. HPC centers typically handle their own user identity management, and those operated by US Government entities must conform to well-specified security policies. Scale of workflow components tends to be more a technical issue; centers typically address these through data transport middleware designed for large scale (e.g. Globus\cite{globus}) and specialized repositories\cite{stansberry2019} which can encapsulate both data transport and storage capabilities. These concerns can also take different forms for commercial cloud compute providers than at leadership-class HPC centers where the ability to scale in compute and data is a goal in itself.

\noindent\textbf{Registries}, as mentioned previously, are places where records of artifacts are stored, and they can be distinct from the repositories that contain the artifacts described by the records. A well-designed registry such as WorkflowHub~\cite{gustafsson2024} is critically important for enabling a FAIR ecosystem anywhere, but particularly as illustrated in~\ref{fig:diagram}, they are crucial. Registries complement repositories by maintaining metadata records for workflow components, ensuring the discoverability and interoperability of research artifacts via metadata and persistent identifiers (PIDs) such as Digital Object Identifiers (DOIs). In this case, the registry stores metadata about the artifacts as well as where they are stored, allowing users to search for artifacts relevant to their needs across repositories both internal and external to the HPC center.

\noindent\textbf{Computing infrastructure}, although not digital, must still be described by rich metadata. Reproducibility and reusability rely on preservation and discovery of the computing contexts in which workflows are executed. It is therefore important that metadata about the resources and infrastructure provided by an HPC center be made available and exportable in useful forms. Tools such as FlowCept \cite{souza2023} help to capture provenance about execution and environment that should be stored and linked to the workflow components. Reproducing execution context is a critical enabler for the seamless cross-center data exchange and workflow portability that IRI needs.

Following the model of the EOSC-Life Collaboratory~\cite{goble_implementing_2021} requires infrastructure for \emph{service} execution, separate from but adjacent to HPC supercomputing resources. Infrastructure provided by OpenShift- and OpenStack-based platforms, for example, enables persistent services for automatically testing workflow component viability as well as continuously indexing and updating registries in response to artifact updates in repositories. Commonly available services running on instances of these platforms also reduce the need for workflows to deploy their own versions (e.g., a key-value store). Validation and benchmarking services help align with NAIRR, ensuring that AI/ML models trained on HPC resources remain FAIR-compliant and reusable even as the FAIR principles are updated in the future.

\noindent\textbf{Authentication and authorization} via Role-Based Access Control (RBAC) mechanisms protect sensitive workflow components while maintaining accessibility for authorized users, ensuring that only the right users can access and use certain components. This is an underappreciated aspect of the ``A'' in FAIR, which stands for Accessibility. Although FAIR is frequently used in conjunction with open science principles, as we noted above FAIR does not strictly imply openness, which is a fact that even seasoned FAIR practitioners can forget.

Rich metadata about workflow components should include information about who can use the components, and this usually falls under the ``R'' for Reusability. When that metadata is made available in the registry, it can feed appropriate authentication and authorization procedures that fall under the ``A'' in FAIR. These problems are not unique to HPC environments, but they must still be solved in FAIR ecosystems for HPC. Authentication and authorization mechanisms, including federated identity management solutions such as OAuth and OpenID Connect, facilitate the secure cross-institutional collaboration necessary for distributed, collaborative computational science.

\noindent\textbf{Metadata standards and interchange formats} are the glue that holds the whole ecosystem together; they play a crucial role in enabling FAIR at all scales. The EOSC-Life Collaboratory's WorkflowHub \cite{gustafsson2024} uses a rich metadata framework for findability with inter-registration integration with sister registries for containers and tools using Application Programming Interfaces (APIs) and shared PIDs. Tools and workflows are marked up with metadata standards from Bioschemas~\cite{gray2017}, encouraging the enrichment of metadata by workflow and tool developers, including licensing and provenance metadata. Abstract Common Workflow Language (CWL)~\cite{crusoe2022} is recommended as a canonical representation of the workflow, and WorkflowHub imports and exports the exchange standard Workflow-RO-Crates~\cite{leo2024}, facilitating automatic builds for each of its entries, regardless of whether the management system natively supports it. The architecture here follows that closely not only because it helps make workflows ``good''~\cite{wcs2024} but also because this architecture seeks to integrate directly with WorkflowHub, in alignment with FASST’s efforts to streamline scientific workflows and training mechanisms.

\section{Encouraging adoption of FAIR among HPC communities}
% DISCUSSION

Encouraging adoption of FAIR practices in HPC environments typically reduces to inducing humans to actively participate. As in many comparable situations within and without computing, a combination of incentives and enforcement mechanisms is likely to be most effective.
Incentives include the creation of reward systems that recognize and incentivize researchers for sharing their components. This can be achieved through citation and attribution mechanisms which formally acknowledge the contributions of those who make their computational resources, data, or code available for public use. At HPC centers, however, there are other currencies available, too, such as increased allocations and job priority. Allocations (measured in node-hours) and job priority (measured in wait time) both represent ways to reward time back to users who actively work to save time for the other users. It will be important to balance such rewards against potential resulting disincentives, such as users' equating reuse of shared components with reduction in resources available to themselves.
%solutions and especially to reuse each other's artifacts.
%disincentivizing users from reusing components which have been shared, due to fears %that they are somehow reducing resources available to themselves.
Receiving allocations of resources at HPC centers, in turn, should become at least partially dependent on FAIR compliance and participation in the workflow component and research artifact sharing ecosystem. Currently, allocations are frequently awarded at no cost to researchers or are governed by proposal processes funded by taxpayers. Artifacts resulting from that usage should (perhaps after some embargo period to protect intellectual investment pending publication) be freely available and reusable. Researchers will want assurances that they will be appropriately credited for their efforts, and so implementation of such mandates will require careful consideration.

\section{Conclusion}
% CONCLUSIONS
%

Fostering FAIR ecosystems for science at scale in HPC centers can make it easier for users of those centers to share and reuse workflow components, in turn reducing duplicated effort and improving scientific productivity. Encouraging adoption through a combination of incentive structures and enforcement mechanisms will further accelerate the integration of FAIR principles in computational science. Ultimately, a well-designed FAIR ecosystem for HPC will not only benefit human researchers but also facilitate the development of autonomous scientific workflows capable of scaling computational research beyond current limitations and perhaps someday even beyond human capabilities.

\begin{acks}
This research used resources of the Oak Ridge Leadership Computing Facility at the Oak Ridge National Laboratory, which is supported by the Office of Science of the U.S. Department of Energy under Contract No. DE-AC05-00OR22725.
\end{acks}

\bibliographystyle{ACM-Reference-Format}
\bibliography{main}

%\appendix

\end{document}